# Expressiveness and machine processability of Knowledge Organization Systems (KOS): an analysis of concepts and relations

**Manolis Peponakis** ; **Anna Mastora** ; **Sarantos Kapidakis** ; **Martin Doerr**

**Abstract**

This study considers the expressiveness (that is the expressive power or expressivity) of different types of Knowledge Organization Systems (KOS) and discusses its potential to be machine-processable in the context of the Semantic Web. For this purpose, the theoretical foundations of KOS are reviewed based on conceptualizations introduced by the Functional Requirements for Subject Authority Data (FRSAD) and the Simple Knowledge Organization System (SKOS); natural language processing techniques are also implemented. Applying a comparative analysis, the dataset comprises a thesaurus (Eurovoc), a subject headings system (LCSH) and a classification scheme (DDC). These are compared with an ontology (CIDOC-CRM) by focusing on how they define and handle concepts and relations. It was observed that LCSH and DDC focus on the formalism of character strings (nomens) rather than on the modelling of semantics; their definition of what constitutes a concept is quite fuzzy, and they comprise a large number of complex concepts. By contrast, thesauri have a coherent definition of what constitutes a concept, and apply a systematic approach to the modelling of relations. Ontologies explicitly define diverse types of relations, and are by their nature machine-processable. The paper concludes that the potential of both the expressiveness and machine processability of each KOS is extensively regulated by its structural rules. It is harder to represent subject headings and classification schemes as semantic networks with nodes and arcs, while thesauri are more suitable for such a representation. In addition, a paradigm shift is revealed which focuses on the modelling of relations between concepts, rather than the concepts themselves.

**Keywords** – Knowledge Organization Systems (KOS); Ontologies; Semantic Web; Computational Linguistics; Expressiveness; Machine Processability

## 1. Introduction

The subject analysis and subject access have always been key issues for managing knowledge and they have been studied by many disciplines and from different perspectives. Library and Information Science (LIS) has a long history in this field and has developed theories and tools within the aforementioned domains. In parallel, Computer Science created the ground for re-examining the field's theory, tools and formalisms. Our study addresses the space between the two aforementioned domains. In this context, Knowledge Organization (KO) theory is considered, embracing perspectives from Philosophy and Cognition. Also, specific types of Knowledge Organization Systems (KOS) are analysed and compared with ontologies in terms of managing concepts and relations as their structural elements.

The aim of this study is twofold. The first aim is to explore and compare the expressiveness (that is the expressive power or expressivity) of each KOS as a subject language, along with comparing this expressiveness against modern schemes of knowledge representation such as ontologies. In this context, language is defined as the vocabulary and its interrelations within a specific KOS regardless of whether it is expressed through words or notation. In the context of Knowledge Organization, expressiveness refers to the ability of KOSs to describe the content of a document, i.e. the way a KOS represents the result of subject analysis. Thus, expressiveness can be defined as "the range of propositions that a certain language is able to express" [1, p.129]. The expressiveness of the system is largely affected by the number of classes and the concepts they contain, as well as the potential relationships between these classes and concepts. In addition, in many cases, the existence of restrictions is essential because "a concept can be determined not only through a conjunction of properties, but also, from time to time, through a disjunction of properties" [2, p.1952].

The second aim of the study is to examine the extent to which the expressiveness of each KOS is machine-processable. Though it is acknowledged that anything that can be computationally managed can be considered machine-processable, for the purpose of this study this notion is regarded in the context of the semantic web, where the effort is in encoding data in such a formalistic way which allows for its management in a predefined, though particularly versatile, way. This structure of information of the "Semantic Web will enable machines to 'comprehend' semantic documents and data, not human speech and writings" [3, p.40]. The term "machine-understandable" is often used as an alternative of machine-processable, yet Antoniou and van Harmelen note that "it is the wrong word because it gives the wrong impression. It is not necessary for intelligent agents to understand information; it is sufficient for them to process information effectively, which sometimes causes people to think the machine really understands" [4, p.3]. By examining the expressiveness of KOSs, evidence is deployed on whether KOSs can be used as a reliable inference tool or whether they should be used simply as a tool for keyword searching.

Two main objectives were identified for delivering the aims indicated. The first objective is to highlight certain features of expressiveness for certain types of KOSs. As regards this objective, the study sets three points of investigation: the features that define expressiveness; the extent to which these features are present in KOSs; and how the expressiveness of KOSs compares to modern systems of knowledge representation, as in ontologies. The second objective is to examine the features of the KOSs' expressiveness that can be transferred and applied to the context of the Semantic Web. For this purpose, two areas are investigated. These are: the conditions under which KOSs are currently being applied to the Semantic Web; and the extent to which the expressiveness of each type of KOS may be applied to the Semantic Web. In order to meet the objectives, two methodological approaches are followed on the basis of implementing comparative analysis. First, a critical literature review of representative types of KOSs and their structure is deployed, commencing from the study of the Knowledge Organization domain within which KOSs are created. Secondly, morphosyntactic analysis is applied to these KOSs in order to elaborate on the types of concepts and relations.

The next sections are organized as follows. Section 2 is the background analysis of the domain in terms of the particular fields discussed above, and hosts two subsections. The first (2.1) deals with concepts in the context of language and cognition while also considering some philosophical aspects. The second subsection (2.2) uses the notion of concepts as defined in the first subsection, along with relations, and outlines their use in Knowledge Organization for generating artificial languages. Section 3 comprises a comparative analysis and critique of the theory of certain KOSs (thesauri, subject headings and classification schemes) as well as an evidence-based analysis involving their linguistic approach. The analysis of the conventional KOSs is performed against a Knowledge Representation system, i.e. an ontology. Finally, Section 4 presents the overall conclusions of this study.

## 2. Background analysis

Knowledge Organization may be perceived mostly through its manifestations, namely in the form of KOSs. However, at an abstract level, Knowledge Organization can also be seen "as the process of structuring of knowledge" [5, p.40]. In order to emphasize on the complexity and the interdisciplinarity of the domain, Hjørland, one of the most influential authors in the field of Knowledge Organization, notes the following: "the relevant literature is very scattered and difficult to synthesize, for it covers, among other fields, philosophy, linguistics, psychology and cognitive science, sociology, computer science, and information science. In addition to the disciplinary scattering of research in semantics, the field is based on different epistemological assumptions whose roots extend back hundreds of years into the history of philosophy" [6, pp.369–370]. In this section, having embraced these remarks, we attempt a brief critique of relevant literature commencing with the importance of concepts in the context of cognition, language and communication, moving towards and analyzing the basic mechanism of a KOS, a determinant of their definition as an artificial language. Subsection 2.2.1 refers to contemporary LIS and the ways it conceptualizes concepts aiming at producing a framework within which the notion of a work's aboutness can be analyzed. Furthermore, there is a presentation of SKOS which is the fundamental mechanism to represent KOSs in the Semantic Web.

## 2.1. Concepts in cognition and language

Acknowledging that the notion of concept is multi-layered, we study concepts in the wider context of cognition and in the context of KOSs based on the principle that "concepts should be considered the building blocks of all forms of KOS" [7, p.38]. According to the Stanford Encyclopedia of Philosophy "concepts are the constituents of thoughts. Consequently, they are crucial to such psychological processes as categorization, inference, memory, learning, and decision making" [8, para. 1]. Lakoff also notices the importance of categories in his classical book entitled "Women, Fire, and Dangerous Things: What Categories Reveal about the Mind" by stating that: "Without the ability to categorize, we could not function at all, either in the physical world or in our social and intellectual lives" [9, p.6]. And, as Rips et al state [10, p.177], "cognitive scientists generally agree that a concept is a mental representation" and "theories in psychology have concentrated on concepts that pick out [a] certain set of entities: categories. That is, concepts *refer*, and what they refer to are categories".

However, as long as concepts are mental representations, they belong to cognition; therefore, it is not easy to observe them directly. Nevertheless, indirect observation is feasible since concepts can be identified within the verbal communication of everyday life through the words they are represented by. In this process, a fundamental problem is the ambiguity of natural language. In many cases, words themselves do not carry a definite meaning, so they cannot be assigned to a specific concept with a one-to-one relationship (polysemy); however, it is the rules of language and a specific context which help us disambiguate. Figure 1 gives a simple example of homonyms, which are also homographs and homophones, and provides an illustration of how a word (in this case "bat") can effortlessly be disambiguated in the context of a sentence. In everyday life the situation is a lot more complicated than this. For example, in the beginning of this section, concepts were referred to as "building blocks". This is a metaphor. Metaphors are very common in human discourse, and the human mind is much trained to manipulate them [11]. The complicated nature of meaningful communication through natural language is colourfully presented by Pinker [12, p.227] who invites us to: "consider how much knowledge of human behavior must be interpolated to understand what *he* means in a simple dialogue like this:

> Woman: I'm leaving you.
> Man: Who is he?"

This example demonstrates the tacit knowledge required in communication and how this is interconnected with a cultural context. As Smiraglia and Heuvel state, "it is in language – representational, traditional, contextual semantics – that 'concept' and 'likeness' clash. It is here that perception becomes a critical process because any given concept is context dependent. And it is here that we return to the semantic hindrance that lies between the expression of thought and the labeling of concepts" [13, p.376].

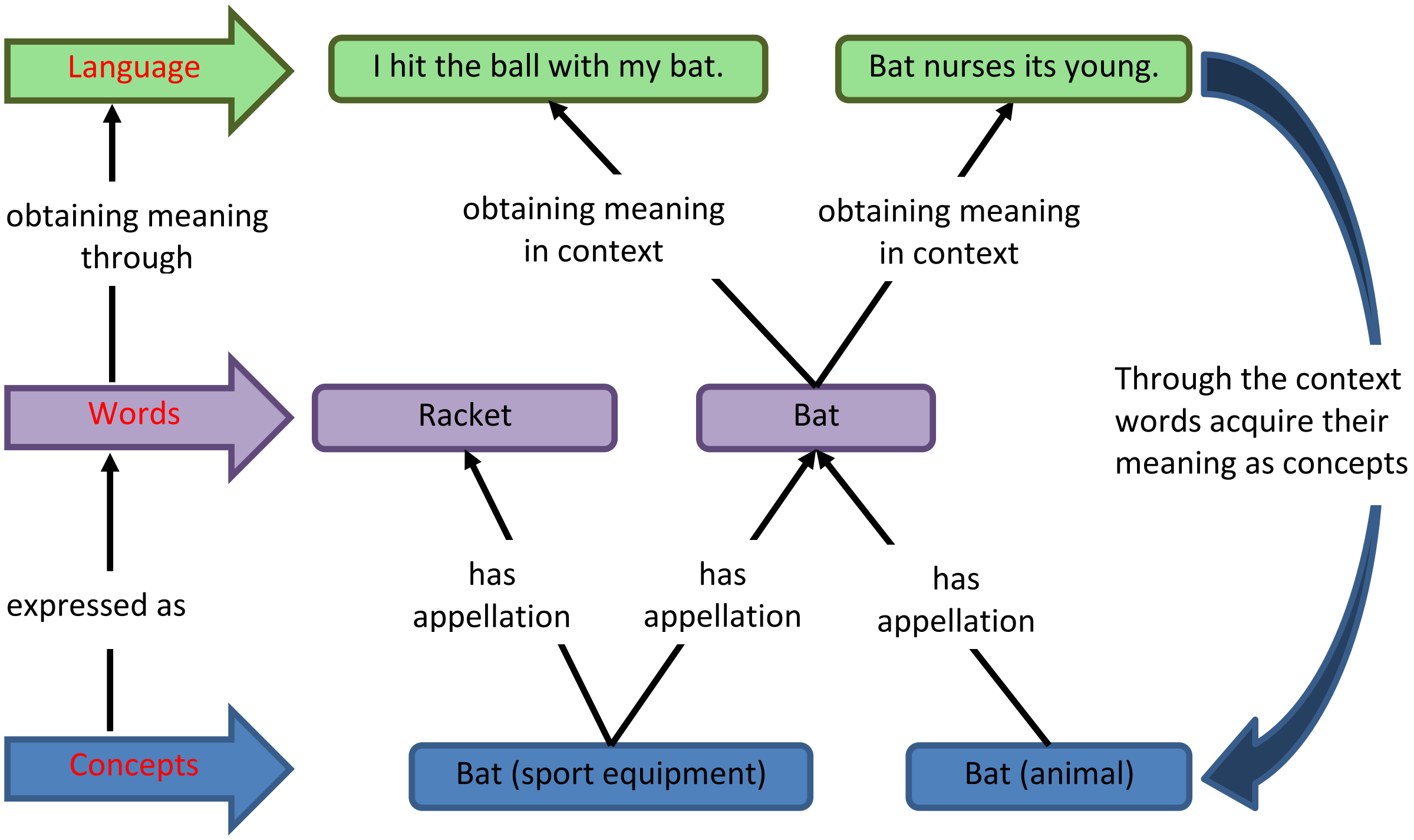

Figure 1: Context and concepts in natural language (based on Wittgenstein's philosophy)

Based on the latter Wittgensteinian philosophy, that is when Wittgenstein "developed a more holistic and pragmatic view of language" [6, p.184] and moved "from the precise logical model of language to the more complex context-dependent model" [14], this study advocates that concepts are the basic units which are expressed through words in natural language and these words obtain their meaning within a certain context. In Figure 1 the basic Wittgensteinian approach is illustrated through an example. We must clarify that the connection between the top and the bottom level of Figure 1 is achieved through the relations established and because of the intermediate levels; not directly. Thus the arrow on the right side of the figure summarizes the result of the relations that connect the levels.

## 2.2. KOS as Artificial Language: Concepts and Relations as Functional Units

All types of KOSs, verbal or not, are based on the ability of the human mind to create and manage categories; KOSs of the LIS domain are no exception. Clearly, not every categorization constitutes a classification [15]; however, categorization is a precondition. Subject access has been a key issue for content providers, having tried hard to develop a range of KOSs to ensure and optimize subject access to their collections. Towards this direction, LIS's KOSs were developed in order to provide subject access to the content of libraries. The KOSs, during their first stages of development in the 19$^{th}$ century, mainly served as “collocating devices” [5, Chapter 4.2.4], while later they were used as indexing tools. Hjørland states that “the most important difference between the different indexing languages […] is between ‘free text systems’ and the others” [16, p.304], therefore he categorizes traditional subject indexing languages into two basic categories; the first contains *Classification systems* which are subdivided to *Enumerative* and *Faceted*, and the second contains *Verbal indexing languages* which are subdivided into *Controlled systems* and *Free text*. The *Controlled systems* are further subdivided into *Precoordinative* and *Postcoordinative* systems [16, fig. 1].

Even though it is rather evident that none of the existing library KOSs reaches the full potential of natural language expressiveness, nonetheless many researchers consider KOSs as languages, albeit “artificial”. Towards this end, Smiraglia and Heuvel state that “from linguistic theory we have learned that both the syntax and semantics are of importance in the understanding of concepts. Since various classifications have been discussed as artificial languages, we are in particular interested [in] the interplay between syntax and semantics in the relations between elementary structures of knowledge from a historical perspective” [13, pp.361–362]. Zavalina states that “the subject representation process is affected by limitations of available subject languages” [17, p.143], while Svenonius propounds that “it is useful to regard a tool for providing subject access to information as a language, since this provides a ready-to-hand conceptual framework that can be used in its analysis and description. This framework characterises a language in terms of its vocabulary, semantics, syntax and pragmatics” [18, p.18]. Based on UDC, Salah et al. suggest that a “complex KOS is rather a language to describe objects in a controlled way than a hierarchical tree” [19, p.56].

Currently, KOSs’ modelling seems to extend in a 2-dimensional space. One dimension is a technical orientation and approach that is hands-on with the transition to the new technological infrastructure. The main focus is given on the structural transformation dealing with issues such as how to convert a given thesaurus to SKOS, how to transcribe SKOS into RDF/XML, and so on. In this way the semantics remain the same and the expressiveness of the new “language” downgrades into the expressiveness of the old one. Again this practice is like converting a txt file to a richer format, such as docx or html, without considering implementing new available features, such as header creation and formatting. What constitutes a paradox is that this approach seems to be welcomed as an effective exploitation of this modern and powerful format! Apparently, using a richer format is a necessary condition; still, not sufficient to provide richer information.

On the other dimension is the theoretical approach which is quite generic. There is a vast and remarkable bibliography dedicated to the epistemological perceptions of classification. Debates are based on modern or postmodern views focusing on aspects such as whether concepts express an objective aspect of reality or not [20]; whether an “interdisciplinary communication […] not grounded in particular disciplines or cultures” [21, p.2247] is feasible; or whether concepts can be understood in the context of specific epistemological theories [22], [23]. Therefore, there is a necessity for moving classification toward a late-modern conception [24]. Olson, mainly driven by feminism studies, attempted a very interesting critique focusing on what underlies the terms and structures of the libraries' KOSs and how these reflect the dominant views of society [25], [26]. This shows that concepts are not neutral representations of an objective world but reflect the marginalization and exclusions adopted within each society. This is why Sperberg-McQueen states that "the assumptions underlying a classification scheme may become effectively invisible and thus no longer subject to challenge or rethinking; for purposes of scholarly work, such invisibility is dangerous and should be avoided" [27, sec. 1].

To conclude, Knowledge Organization Systems, as a way of modelling knowledge, play a key role in the creation of new knowledge because the way we model our understanding both reflects and affects our ability to understand. "In other words, how knowledge is organized and represented depends largely on the understanding of how knowledge is generated and realized" [28, p.364]. This kind of dependency has been revealed in a study which concluded that the need for the ontological formalism of OWL leads ontology producers to "create" literals which do not exist in the specialized dictionaries of a certain domain's vocabulary [29]. In addition, the modelling of knowledge does not only affect our thinking patterns but also, under certain conditions, can monitor the correctness of our assumptions. Goguen names this process as "'experimental philosophy' in that it asks for experiments to be performed in testing the correctness of hypotheses about language meanings" [30, p.546].

Admitting that concepts are the building blocks of all forms of KOSs entails that relations must be the glue between them, since having no strong connections between the blocks makes no solid building. The majority of KOSs are governed by equivalence relationships (basically in terms of managing synonyms or quasi-synonyms), hierarchical relationships and associative ones; all of them often go by the name Paradigmatic relationships. Associative relationships constitute a very generic and ambiguous category since "any paradigmatic inter-term relationship that is not hierarchical, nor of equivalence, is nowadays usually described as "associative"" [31]. On the other hand, there are Syntagmatic relationships which involve lexical relations (often intra-term). Green states that Paradigmatic relationships are a closed class, while Syntagmatic relationships are an open class [32], which allows for "the expression of new, complex meanings", as Engerer states [33, p.1485].

In the following subsections two cases are presented where KOSs are handled as artificial languages.

### 2.2.1. Conceptualizing the KOS: FRSAD

Given the overall influence of the Functional Requirements for Bibliographic Records (FRBR), on which the Resource Description and Access (RDA) standard has been based, a study focusing on the modelling of subjects would be incomplete without considering the Functional Requirements for Subject Authority Data (FRSAD) [34]. An exhaustive analysis or critique of FRSAD is beyond the scope of this study, yet, a few aspects are addressed in order to put this work in the context of modelling *aboutness* according to FRSAD.

IFLA decided to extend the principles of FRBR to subjects. To fulfil this need, the "Working Group on the Functional Requirements for Subject Authority Records (FRSAR)" delivered the "Functional Requirements for Subject Authority Data (FRSAD): A Conceptual Model". The basic rationale of FRSAD is based on the precondition that a *work* has a *thema* as a subject which has a *nomen* as an appellation. "Any entity used as a subject of a *work*" is defined as *thema* and "any sign or sequence of signs (alphanumeric characters, symbols, sound, etc.) that a *thema* is known by, referred to, or addressed as" is defined as a *nomen* [34, p.15]. In essence, a *thema* functions as a wrapper element to which all concepts of all kinds of KOSs can be assigned. Additionally, FRSAD defines the relationships between the *thema* and the *nomen*, between a *nomen* and another *nomen*, as well as the relationships between a *thema* and another *thema*. The latter is subdivided into Hierarchical Relationships, the Generic Relationship, the Whole-Part Relationship, the Instance relationship, the Polyhierarchical Relationship, other Hierarchical Relationships and the Associative Relationships, which cover affiliations between pairs of *themas* that are not related hierarchically. Such types of relationships are Cause – Effect, Process – Agent, and Raw material – Product.

Being an abstract model, FRSAD does not provide certain specifications for applying the aforementioned relationships to certain KOSs, and it does not provide a specific structure for implementation. Yet, the attempted abstraction, which was based on existing types of KOSs used by libraries, raises some concerns and FRSAD has been the subject of multilateral critique. Furner points to the fact that, since all FRBR entities could serve as subjects, the "FRBR model amounts to an ontology of subjects" [35, p.500]; under this condition, FRSAD becomes a superset of FRBR which constitutes a paradox. This overlapping of the third group (i.e. the entities of *concept, object, even*t and *place*) with the other two groups (i.e. the entities of *work, expression, manifestation* and *item* in group 1 and *Person, Corporate Body* in group 2) is also mentioned by Gemberling who states that the third group is the only non-exclusive (disjointed) group in relation to the other two groups [36, p.446]. In addition, Panzer questions: "Is *thema* meant to be skos:Concept or owl:Thing? What exactly *does* subject authority data represent: the world of subjects, or subjects in the world?" [37, p.167]. Panzer's question is essential since a SKOS concept is, de jure, an instance of an OWL class.

FRSAD tries to model *aboutness*, which "generally has been used to refer to the relationship between a document and its subject" [38, p.2471], without making a clear distinction whether *aboutness* "should be conceived not as a property of works but rather as a relation, constructed by a particular person at a particular time, between a particular set of works and a particular linguistic expression (i.e., a name or label*)*" or whether "subjects are real things that exist separately from the linguistic expressions that we use to name them, and that it is possible to determine 'the' subject(s) of any given work" [34, pp.10–11]. From our point of view, it is important to clarify that the *thema* is not a property of the resource but a property of a specific KOS that can be assigned to (the description of) the resource. For example, when considering the work "Birds of Africa" using the LC subject headings for cataloguing, the one and only *thema* could be the one that is represented by the heading "Birds -- Africa" (alternative nomen: URI http://lccn.loc.gov/sh2009117134). If the Eurovoc Thesaurus is used, instead of LCSH, then the result would be "Bird" (alternative nomen: URI http://eurovoc.europa.eu/5760) and "Africa" (alternative nomen: URI http://eurovoc.europa.eu/281), i.e. two different *themas*.

An essential dichotomy issued by FRSAD is the distinction of a nomen into two types: the identifier and the controlled name. Previously, KOSs that implemented verbal indexing were term-centric in a way that the concept was the outcome of the words representing it; thus, terminological aspects played a key role. To elaborate on the previous statement, in traditional implementations of access points, *nomens* had a twofold role, being used both as names and as identifiers in a way that they formed a unique string assigned to the entity in order to disambiguate. The ultimate purpose was to achieve string matching during the information retrieval process. While this name is used as a product of natural language, by its nature it also carries the semantics of natural language and expresses a social discourse [26]. The dominance of computers emerged with the capability to use random identifiers which do not carry any of the name features. From ID numbers to ORCID and from DOI to Record Numbers there is a vast diversity of such identifiers.

The distinction between these different types (i.e. identifier and controlled name) is essential. To indicate the significance of this distinction the LC authority record referred by the URI http://lccn.loc.gov/n81038515 will be used. According to LC, this entity has preferred label value "Macedonia (Republic)". Until very recently, Greek governments had been declining, wherever applicable, the use of this name and had been referring to the country as Former Yugoslav Republic of Macedonia (FYROM). This had not been a dispute over the existence of the country or its sovereignty; only over its name. Neither the previously mentioned URI nor the entity itself had triggered any dispute; although the name "Macedonia (Republic)" and its use could have easily grown to a diplomatic incident. Matthew Nimetz, the United Nations Special Representative for this dispute, had been trying for more than two decades to find a compromising solution [39]. In late 2018, after 27 years of negotiations as well as many protests from both sides against the proposed names, the two countries finally agreed in using the name North Macedonia. Another typical example of the importance of the label is the use of the terms "Anno Domini" (abbreviation AD) and "Current Era" or "Christian Era" (abbreviation CE). The usage of one or the other has raised issues of political correctness[1] since "naming is a vital issue for human cultures. Names are not random sequences of characters or sounds that stand just as identifiers for the entities, but they also have socio-cultural meanings and interpretations" [40, p.33]. Therefore, identifiers and natural language names have different properties, and call for different modelling.

[1] For an analysis of these issues see: http://en.wikipedia.org/wiki/Common_Era and https://en.wikipedia.org/wiki/Anno_Domini

It should be noted that it is difficult to talk about the *thema* without using one of its *nomens*, in the same way it is acknowledged that "it is necessarily difficult to talk of a concept without using at least one term" [41, p.179]. Irrespectively of whether we are able to discuss the *thema* with or without referring to a *nomen*, the definition of *thema* is a fundamental issue. Concluding this subsection and according to FRSAD's[2] terminology, our work studies the nature of the *thema* itself by referring to its paradigmatic relationships, and approaches it by implementing computational linguistics on its *nomen*. This aspect is viewed both in terms of KOSs' theory, and in terms of language rules and norms which help for identifying certain patterns. As it happens with patterns [43, p.579], they assist in making assumptions which, in this case, are oriented towards designating the nature of concepts and relations within KOSs regarding the role they hold in terms of the KOSs expressiveness.

### 2.2.2. Representing the KOS: Simple Knowledge Organization System (SKOS)

SKOS is a W3C recommendation designed for the representation of structured controlled vocabularies such as thesauri, classification schemes, and subject-heading systems. It aims to structure data in order to be machine-processable in the environment of the semantic web and linked data. As Shiri puts it: SKOS "aims to build a bridge between the world of knowledge organization systems [...] and the linked data community, with the goal of bringing benefits to both" [44, p.18].

According to W3C "The SKOS data model views a knowledge organization system as a concept scheme comprising a set of concepts. These SKOS concept schemes and SKOS concepts are identified by URIs, enabling anyone to refer to them unambiguously from any context, and making them a part of the World Wide Web" [45]. SKOS defines a concept as "an idea or notion; a unit of thought. However, what constitutes a unit of thought is subjective, and this definition is meant to be suggestive, rather than restrictive" [45]. Furthermore, "a skos:Concept is intended to provide a neutral target for migrating a wide diversity of KOS concepts to the Web of Data" [46, p.39] and does not force the modeller to distinguish between classes and individuals [47, p.87]. Thus SKOS does not determine the nature of concepts. On the other hand, SKOS does define explicitly and formalistically the relations, i.e. the properties, of concepts. Therefore, relations are limited to associative and hierarchical ones, such as those defined in a thesaurus; notwithstanding more precision by using extensions like "skos:broaderTransitive". Additionally, SKOS provides relations for mapping concepts of different concept schemes using properties such as "skos:closeMatch".

---

[2]FRSAD is incorporated in the consolidate edition of FRBR, known as Library Reference Model (LRM) [42], which was endorsed by the IFLA Professional Committee in the summer of 2017. But, in the LRM there are no *concept*, *object* and *event* entities of FRSAD; neither the term *thema* which is generalized and renamed as *res* (Latin for "thing") in order to serve as the top entity in the hierarchy. Basically, LRM is not concerned with Aboutness, and it ostracizes the structure of KOSs from the issues of concern. The only definition that LRM contains is that a *res* has *nomens* and that it may be related with another *res*.

It would be useful to note that KOSs are schemes for organizing knowledge while SKOS is used for expressing KOSs in the context of the semantic web. Indeed, due to the fact that SKOS concepts are based primarily on URIs and less on natural language labels, there are no limits deriving from the expressiveness of natural language. This makes it possible for us to refer to things not by their names but using a URI. Finally, it should be noted that SKOS allows for flexibility in the wording of URIs using properties such as “skos:hiddenLabel”. SKOS has a significant limitation which is incompatible with LIS practice. It does not allow the names of entities to be modelled, i.e. since they are RDF literals, there is no option to establish relationships between the appellations of the entities. This decision was made “in order to keep the main SKOS specification as simple as possible” [46, p.42]. Still, modelling the entity name is important, thus, the “SKOS eXtension for Labels” [48] has emerged to fill the gap by allowing relationships between nomens.

To conclude this section, it should be noted that the transition to and compliance with the principles of SKOS bring to surface some latent issues of KOSs which were not previously detected due to the use patterns of KOSs [49]. This is mainly because the transition of KOSs to SKOS is mostly implemented at the structural level. Particularly for the transition of thesauri, Kless et al critique the methods used for reengineering thesauri as they do not make use of the potential of formal languages such as OWL in terms of managing semantics for detecting logical errors and improving vocabularies [50]. The aforementioned perception and practice seems not only well established within the community responsible for the transition of KOSs but also for a large amount of data which is transitioned to linked data [51].

# 3. Analysis of representative KOSs

After having presented the theoretical background of this study, this section analyzes concepts and relations within specific KOSs. The analysis is based both on the theory of KOSs and relevant literature as well as on evidence generated by implementing Natural Language Processing (NLP) techniques, namely Part of Speech (POS) tagging and morphosyntactic analysis. As mentioned in the Introduction of this paper, the NLP was implemented in order to provide evidence-based data about the linguistic characteristics which affect both the expressiveness of KOSs and the effectiveness of their transition in the context of the Semantic Web. In this context, two kinds of comparison are involved in this study. First is the comparison among certain conventional KOSs. Then, these KOSs are compared against an ontology. Specifications of the scope and overall process are given in subsection 3.1, while subsection 3.2 presents evidence from the linguistic analysis. In subsection 3.3 the comparative analysis is deployed, and in subsection 3.4 the critique concerning each KOS as well as the ontology is presented.

At this point it should be emphasized that the purpose of this analysis is neither to disprove the suitability of KOSs in terms of their initial purpose of creation nor to question their use in information systems of the 20th century and since then. What is examined is the suitability of their transition to the Linked Data environment, as already happens in all three examined KOSs which are available as Linked Data, alongside their use as final products in the context of the Semantic Web.

## 3.1. Methodology and datasets

In order to fulfil the aim of this study, certain KOSs were examined through literature with a special focus on their structure. For the intra-term relations, as well as for the ontology's properties-relations, natural language processing was implemented. The KOSs are representative of those most used by libraries. These are a subject headings system, i.e. the Library of Congress Subject Headings (LCSH), a thesaurus, i.e. the multilingual Eurovoc, a classification system, i.e. the Dewey Decimal Classification (DDC) and an ontology, namely the CIDOC-CRM. The latter was used for examining the approach to knowledge organization and representation that is relatively new to the ecosystem of libraries.

Terms in KOSs comprise keywords or short phrases, i.e. elliptical sentences. This means that they do not offer much contextual information which is important for morphosyntactic analysis. For example, if the word "search" is provided out of any additional context, it cannot be morphosyntactically identified unambiguously; it could be either a noun or a verb. On the contrary, in the sentence "Yesterday I performed a successful search", the word "search" is recognized as a noun, while in the sentence "Search for the tickets", the same word is recognized as a verb. This is a problem when implementing morphosyntactic analysis, both for humans and algorithmic processes. It should be emphasized though that the aforementioned ambiguousness is found to a varying extent among languages. For example, the English language, like Vietnamese, is not highly inflectional so, when the context of an English word is limited, it is likely that a number of parts-of-speech will not be properly designated. In other languages such as Greek, Finnish, Hungarian, Turkish and Russian, due to their rich morphology and high inflectionality, the morphosyntactic analysis is performed more accurately with limited contextual information. Information about the morphological typology of languages is given by Comrie [52]; more specifically, information about inflectional morphology is found in the work of Stump [53]. Additionally, the work of Pirkola [54] highlights what the morphological typology of languages, i.e. the shared or different linguistic features, means for Information Retrieval.

In our study, taking into account these characteristics of languages, the Greek versions of the aforementioned KOSs were used since Greek provides more solid ground for the POS tagging of elliptical sentences. It should be mentioned, though, that even in Greek there are some cases where POS tagging is challenging in terms of the syntactic ambiguity caused by the morphology of words. An example is the case of the plural number of feminine nouns in the nominative or accusative case which could also be annotated as the second person of verbs in the singular. Such is the case with "υποδιαιρέσεις" (ypodiairéseis), which stands for either *subdivisions* or *subdivide*. Yet, in the context of KOSs, such cases are rare, since few verbal types of POS have been allocated during post-processing.

More specifically, the EU's multilingual thesaurus Eurovoc was used (version 4.4), the Greek version of which comprised 6,883 descriptors. In terms of LCSH, the authorities version of the National Library of Greece (NLG) was used which contains the Greek translation of selected LC Subject Headings. From 55,204 subject headings in total we isolated the ones that were specifically declared to be a translation of LCSH, numbering 19,986 records. All MARC21 subfields other than $a and $x were dismissed as they either comprised Geographic subdivisions ($z), i.e. Named Entities, Chronological Subdivisions ($y), or carried information that was not within the scope of this study. The entries we finally used were at the subfield level, meaning that every $a and $x formed an individual entry. This led to 30,073 entries. After having removed $a and $x as values themselves, we then eliminated a number of encoding and punctuation problems, and finally ensured there was no duplication of subfields. The number of uniquely identified subfields reached 10,308. The subfield level was chosen for processing this data because subfields are considered the smallest autonomous functional units of a heading. Concerning DDC, the 13th abridged edition translated into Greek by the Greek National Documentation Centre was used. It should be noted that the Greek version embodies the translated 21st American edition for the domains of Greek literature and history as well as the history of Cyprus. DDC delivered 5,398 entries, which were reduced to 3,811 after the deduplication process. In all cases, further processing was necessary in order to handle the data in a uniform way. Finally, the Greek translation of the original 4.1 version of the CIDOC Conceptual Reference Model (CIDOC-CRM) was used. For the purpose of this study, Classes and Properties were treated separately.

Each examined KOS comprised an autonomous set, i.e. it was not a subset of a larger dataset. Despite the fact that the NLG corpus derives from LCSH, it does not count as a subset of LCSH as it does not deliberately exclude any part of the original set. Its limitation is that it concerns the Greek publishing distribution and, therefore, its enrichment is adjusted accordingly. In essence it is a representative sample, since it hosts headings from all domains. The Greek version of the DDC follows the respective English version, with just a necessary extension for the domain of Greek history. The version of Eurovoc is the full Greek version, i.e. comprising all the descriptors, and CIDOC-CRM is a one-to-one translation of the original. In addition, we have to mention that, for the purpose of this study, only the established form of each concept was used; thus, non-preferred terms, scope notes and any other record elements were excluded.

The software used for the Part of Speech (POS) tagging has been developed by the Greek Institute for Language and Speech Processing and it is part of a Natural Language Processing (NLP) suite which is based on both machine learning algorithms and rule-based approaches [55]. The software is freely available as a web service at http://nlp.ilsp.gr/soaplab2-axis/. At this point it should be mentioned that this tool is a third-party resource which is used as an end product in the context of this study. The rationale behind this choice was that the tool was developed by an official research institution which is dedicated to the study of the Greek language, both oral and written, as well as to the development of tools for computational processing.

By and large, we followed a common way for data pre-processing. We created a plain text file with the required elements from each KOS, which was then transformed to an xces xml file in order to submit it to the tagger. The output of the service was a POS tagged xces xml file which was post-processed in order to be merged with the initial file and enriched with additional information required for the data analysis. All datasets were finalized as delimited csv files. Additionally, some symbols (!, % and *) were removed in order to process the data more effectively. All metrics refer to the number of unique entries per KOS, meaning that duplicate values were removed during pre-processing. In the deduplication process, capital letters were not considered equal to their corresponding small letters, so "Freedom" and "freedom" were recognized as two different instances. Concerning the deduplication process according to the non-elimination policy of small and capital letters, the decision was based on the implementation of a per-KOS analysis which led us to interpret that, originally, a variation in the use of small or capital letters was purposive.

Based on relevant bibliography [56, 57], we decided upon the parts of speech that we further analyzed according to whether they could be considered entities or relations. Therefore, we considered as entities the Nouns (common and proper), the Abbreviations which comprised capital letters only, specific Pronouns, i.e. demonstrative, personal and relative indefinite, and the Foreign Words, which mainly comprised of either common nouns or Named Entities. Regarding the unveiling of latent relations within the studied KOSs, we observed that, although verbs did not have a strong presence, other ways of expressing some kind of relation between concepts were there. A significant part were "Adpositions" comprising words like *from*, *for*, *in, to*, *by*, *based*, on, against, due and *with*. Some examples of such cases are *Books for children*, *Machinery in industry*, *Crimes against peace*, *Discriminations against women based on sex.* Therefore, we defined Verbs, Conjunctions, Adpositions, Negative Particles, Comparative and Superlative Adjectives, Adverbs and Possessive Pronouns as relations. The remaining identified parts of speech were not included in one of these categories; therefore they are not depicted in the results. This omission also explains why certain figures in Table 1 do not account for 100% of the respective sum totals. A detailed list of the parts of speech that the ILSP-NLP POS tagger identifies is available at http://nlp.ilsp.gr/nlp/tagset_examples/tagset_en/.

## 3.2. Cumulative results

In this subsection we present the metrics resulting from the Part-of-Speech tagging procedure. In order for the reader to better comprehend the results in the context of this study, the definitions of "entries", "tokens" and "words" are given. An *entry* is each alphanumeric sequence of characters or groups of characters defined as an individual listing in the contents of a KOS. To elaborate on this definition, an entry is any descriptor of the thesaurus, any subfield of LCSH, and any lexical equivalent to each DDC notation. For example, *Coaches (Athletics)* is an entry from LCSH. A *Token* is a significant unit (character or group of characters) that was assigned a POS characterization, whether it is a word, digit(s) or punctuation mark. The entry *Coaches (Athletics)* comprises four tokens, i.e. two nouns and two parentheses. *Words* are the remaining sequences of characters if we exclude digits and all punctuation marks as well as other symbols from the identified tokens; therefore, the aforementioned example counts as two words.

Table 1 is categorized per KOS and depicts figures as well as average amounts and percentages depending on the nature of each metric. The three aforementioned data categories (entries, tokens and words) are depicted in the first part (first five rows) of Table 1. Additionally, in the first part of Table 1, we also present the average number of words within each entry per KOS. The second part of Table 1 consists of the POS categorization of entities and relations per KOS. The column “POS no” depicts the number of words tagged for the respective part of speech; the column “POS no/ entries (avg)” lists the average number of words in each POS against the total number of entries in each KOS; finally, the column “POS no/ words (%)” depicts each POS as a percentage of the total number of words in each KOS.

The majority of words were categorized into either entities or relations. Specifically, according to Table 1 the overall rate of categorized words is 68.29% in Eurovoc, 73.14% in LCSH (NLG corpus), 67.42% in DDC, 84.5% in CIDOC-CRM classes and 87.20% of CIDOC-CRM properties. This categorization does not comprise Adjectives (basic) which are distributed as follows: Eurovoc 3,678 (24.41%), LCSH (NLG corpus) 4,929 (24.05%), DDC 3,179 (21.75%), CIDOC-CRM classes 20 (15.50%) and CIDOC-CRM properties 23 (4.20%). These rates, added to the POS distribution rates of Table 1, cover the analysis of 92.70% of Eurovoc words, 97.19% of LCSH (NLG corpus) words, 89.17% of DDC words, 100% of CIDOC-CRM class words and 91.41% of CIDOC-CRM properties words.

The data depicted in Table 1 provides only part of the evidence needed in order to examine the latent relations which are governed by linguistic rules, i.e. morphosyntactic patterns, in KOSs. Observable syntactic and semantic patterns (as compared to patterns that rarely or never occur in natural language) are a manifestation of human language processing and may serve as the basis for inductions about mental processing [58, p.1568]. This kind of investigation has various applications. Lioma and Ounis [59] dealt with query reformulation and recommended a technique called syntactically-based query reformulation (SQR) based on the association of the Lexicon with the respective shallow syntactic evidence of the lexical representation of queries. Although trends of the time suggested that, in natural languages, words are the carriers of meaning while sentences are just shallow syntactic formations of parts of speech, they advocated that syntactic information (that is, the categorization according to parts of speech) could be exploited for query reformulation, particularly in the case of textual information retrieval. More specifically, they claim that taking into account the syntax may increase the potential for retrieving more relevant items, since it plays an important role (though indirect) in carrying meaning by defining the relations which associate the words in coherent sentences. In brief, they accept that shallow syntactic information can model language structure, revealing those structures which are more likely to occur or/and co-occur. More specifically, shallow syntactic structures are sentential fragments, the lexical entities of which have been replaced by their respective shallow syntactic category in order to form the so-called part-of-speech blocks (POS blocks). In summary, they confirm that the high frequency of shallow syntactic fragments correspond to lexical units which carry meaning. And, although some languages may share limited characteristics, nevertheless, according to Murphy, languages do tend to have similar morphosyntactic properties [60, p.435].

Table 2 lists, in descending order, the twenty most frequent occurrences concerning part of speech combinations, i.e. syntactic patterns, for each analyzed KOS and the ontology. The percentage of occurrences of each combination appears in the parentheses. In addition, the analysis of patterns reveals the distribution of the unique syntactic patterns per KOS. In detail, Eurovoc has 286 unique POS combinations (which gives an average of 24.06 entries per pattern), LCSH (NLG corpus) 474 (which is 21.75 entries per pattern on average), and DDC has 1,128 (an average of 3.38 entries per pattern), while CIDOC-CRM has 8 classes (an average of 10.13 entries per pattern) and 41 properties (an average of 5.44 entries per pattern). As it turns out, POS patterns in Eurovoc seem to be more consistent in form; in LCSH (NLG corpus) syntactic patterns are somewhat looser, while in DDC repeated syntactic patterns hold a rather small percentage. This observation confirms to a certain extent that formalism upon the lexical representations in verbal indexing leads to a decrease in the variety of patterns.

In terms of the parts of speech that were categorized as *relations*, appearing in Table 1, further analysis shows that, with the exception of CIDOC-CRM properties, very few of the aforementioned parts of speech are allocated in the twenty most frequent syntactic patterns. More specifically, Verbs are present only in CIDOC-CRM properties, occurring in all twenty patterns. Conjunctions appear in two patterns in LCSH (NLG corpus), in five patterns in DDC and in two patterns in CIDOC-CRM properties; CIDOC-CRM classes and Eurovoc do not host conjunctions in their most frequent patterns. Adpositions appear in three Eurovoc patterns, in two LCSH (NLG corpus) as well as in DDC, and seven in CIDOC-CRM properties. Adverbs appear only once in each KOS and in CIDOC-CRM classes, while in CIDOC-CRM properties two patterns with adverbs are recorded. The two remaining parts of speech, i.e. particle negatives and possessive pronouns, do not appear in the twenty most frequent syntactic patterns.

There follows a more detailed and comparative analysis concerning the expressiveness of KOSs.

| | Eurovoc | | | LCSH (NLG corpus) | | | DDC | | | CIDOC (classes) | | | CIDOC (properties) | | |
|---|---|---|---|---|---|---|---|---|---|---|---|---|---|---|---|
| Number of entries | 6882 | | | 10308 | | | 3811 | | | 81 | | | 223 | | |
| - Number of tokens | 15234 | | | 23670 | | | 16414 | | | 129 | | | 552 | | |
| - Number of words | 15067 | | | 20497 | | | 14613 | | | 129 | | | 547 | | |
| - Words per entry (avg) | 2.19 | | | 1.99 | | | 3.83 | | | 1.59 | | | 2.45 | | |
| | POS no | POS no/ entries (avg) | POS no/ words (%) | POS no | POS no/ entries (avg) | POS no/ words (%) | POS no | POS no/ entries (avg) | POS no/ words (%) | POS no | POS no/ entries (avg) | POS no/ words (%) | POS no | POS no/ entries (avg) | POS no/ words (%) |
| **POS defined as Concepts/Entities (sum)** | **9652** | **1.40** | **64.06%** | **13321** | **1.29** | **64.99%** | **7552** | **1.98** | **51.68%** | **108** | **1.33** | **83.72%** | **112** | **0.50** | **20.48%** |
| - Nouns Common | 8443 | 1.23 | 56.04% | 11825 | 1.15 | 57.69% | 6443 | 1.69 | 44.09% | 105 | 1.30 | 81.40% | 111 | 0.50 | 20.29% |
| - Nouns Proper | 653 | 0.09 | 4.33% | 1209 | 0.12 | 5.90% | 997 | 0.26 | 6.82% | 1 | 0.01 | 0.78% | 0 | 0.00 | 0.00% |
| - Abbreviations | 240 | 0.03 | 1.59% | 11 | 0.00 | 0.05% | 5 | 0.00 | 0.03% | 1 | 0.01 | 0.78% | 0 | 0.00 | 0.00% |
| - Pronouns[3] | 2 | 0.00 | 0.01% | 0 | 0.00 | 0.00% | 12 | 0.00 | 0.08% | 0 | 0.00 | 0.00% | 0 | 0.00 | 0.00% |
| - Residual, foreign word | 314 | 0.05 | 2.08% | 276 | 0.03 | 1.35% | 95 | 0.02 | 0.65% | 1 | 0.01 | 0.78% | 1 | 0.00 | 0.18% |
| **POS defined as Relations (sum)** | **637** | **0.09** | **4.23%** | **1671** | **0.16** | **8.15%** | **2300** | **0.60** | **15.74%** | **1** | **0.01** | **0.78%** | **365** | **1.64** | **66.73%** |
| - Verbs | 18 | 0.00 | 0.12% | 34 | 0.00 | 0.17% | 166 | 0.04 | 1.14% | 0 | 0.00 | 0.00% | 234 | 1.05 | 42.78% |
| - Conjunctions | 98 | 0.01 | 0.75% | 614 | 0.06 | 3.00% | 970 | 0.25 | 6.64% | 0 | 0.00 | 0.00% | 14 | 0.06 | 2.56% |
| - Adpositions | 368 | 0.05 | 2.44% | 726 | 0.07 | 3.54% | 796 | 0.21 | 2.44% | 0 | 0.00 | 0.00% | 93 | 0.42 | 17.00% |
| - Particles Negative | 35 | 0.01 | 0.23% | 17 | 0.00 | 0.08% | 30 | 0.01 | 0.21% | 0 | 0.00 | 0.00% | 0 | 0.00 | 0.00% |
| - Adjectives[4] | 16 | 0.00 | 0.11% | 22 | 0.00 | 0.11% | 83 | 0.02 | 0.57% | 0 | 0.00 | 0.00% | 6 | 0.03 | 1.10% |
| - Adverbs | 82 | 0.01 | 0.54% | 213 | 0.02 | 1.40% | 220 | 0.06 | 1.51% | 1 | 0.01 | 0.78% | 17 | 0.08 | 3.11% |
| - Pronouns possessive | 5 | 0.00 | 0.03% | 27 | 0.00 | 0.13% | 12 | 0.00 | 0.08% | 0 | 0.00 | 0.00% | 1 | 0.00 | 0.18% |

Table 1: Metrics of Part of Speech Tagging for each KOS

[3] Relative Indefinite, Personal and Demonstrative

[4] Comparative and Superlative

| Eurovoc | | LCSH (NLG corpus) | | DDC | | CIDOC (Classes) | | CIDOC (Properties) | |
|---|---|---|---|---|---|---|---|---|---|
| POS pattern | Total No and rate | POS pattern | Total No and rate | POS pattern | Total No and rate | POS pattern | Total No and rate | POS pattern | Total No and rate |
| Adj + N | 2182 (31.71%) | N | 3353 (32.53%) | N | 634 (16.64%) | N | 38 (46.91%) | V + Adp | 59 (26.46%) |
| N | 1364 (19.82%) | Adj + N | 1936 (18.78%) | Adj + N | 387 (10.15%) | N + N | 23 (28.40%) | V | 40 (17.94%) |
| N + N | 819 (11.90%) | N + Punct + Adj | 677 (6.57%) | N + N | 156 (4.09%) | Adj + N | 13 (16.05%) | V + N + Art | 21 (9.42%) |
| N + Art + N | 454 (6.60%) | N + N | 548 (5.32%) | Dig | 112 (2.94%) | N + Adj + N | 2 (2.47%) | V + N | 17 (7.62%) |
| Res | 191 (2.78%) | N + OPunct + N + CPunct | 369 (3.58%) | N + Conj + N | 112 (2.94%) | Adj + Adj + N | 2 (2.47%) | V + N + Adp | 11 (4.93%) |
| N + Adj + N | 171 (2.48%) | N + Adp + N | 361 (3.50%) | Adj | 63 (1.65%) | N + Res + Abr | 1 (1.23%) | V + Adj + N | 8 (3.59%) |
| Adj + Adj + N | 143 (2.08%) | N + Conj + N | 311 (3.02%) | N + Adj + N | 61 (1.60%) | N + Adj | 1 (1.23%) | V + N + N | 6 (2.69%) |
| N + Adp + N | 118 (1.71%) | Adj | 261 (2.53%) | Adj + Adj + N | 50 (1.31%) | Adv + N | 1 (1.23%) | V + Conj + V + N + Art | 5 (2.24%) |
| Adj | 103 (1.50%) | Adj + Adj | 134 (1.30%) | N + Art + N | 44 (1.15%) | - | - | V + Adj + N + Art | 5 (2.24%) |
| Adj + N + N | 93 (1.35%) | Adj + N + Punct + Adj | 124 (1.20%) | Adj + N + Conj + N | 40 (1.05%) | - | - | V + Conj + V + N | 4 (1.79%) |
| N + Art + Adj + N | 68 (0.99%) | N + Punct + N | 120 (1.16%) | Adj + Conj + Adj + N | 37 (0.97%) | - | - | V + Art + N + Adp | 4 (1.79%) |
| Adj + Adj | 59 (0.86%) | N + Art + N | 108 (1.05%) | N + Punct + N + Punct + N | 33 (0.87%) | - | - | V + Adv + N | 4 (1.79%) |
| N + Adp + Art + N | 42 (0.61%) | N + Adj + N | 85 (0.82%) | Adv + N | 31 (0.81%) | - | - | V + Adj + Adj | 4 (1.79%) |
| N + Art + Abr | 42 (0.61%) | N + OPunct + Adj + N + CPunct | 77 (0.75%) | N + Conj + Adj + N | 31 (0.81%) | - | - | N + V + Adp | 3 (1.35%) |
| N + N + N | 39 (0.57%) | N + Res | 72 (0.70%) | Adj + N + N | 28 (0.73%) | - | - | V + V + Adp | 2 (0.90%) |
| Adj + N + Adj + N | 37 (0.54%) | Adv + N | 67 (0.65%) | N + Conj + N + N | 28 (0.73%) | - | - | V + Art + N + Art | 2 (0.90%) |
| Adv + N | 37 (0.54%) | N + Punct + N + Art | 60 (0.58%) | Adj + Adj | 27 (0.71%) | - | - | V + Adv | 2 (0.90%) |
| Adj + N + OPunct + Abr + CPunct | 36 (0.52%) | Adj + Adj + N | 48 (0.47%) | N + Adp + Adj + N | 27 (0.71%) | - | - | V + Adp + N | 2 (0.90%) |
| N + Adp + Adj + N | 36 (0.52%) | Adj + N + Adp + N | 48 (0.47%) | N + Adp + N | 27 (0.71%) | - | - | V + Adj + Adp | 2 (0.90%) |
| N + Res | 34 (0.49%) | Adj + N + Conj + N | 45 (0.44%) | Res | 26 (0.68%) | - | - | V + V | 1 (0.45%) |
| **Sum of 20** | **6068 (88.17%)** | **Sum of 20** | **8804 (85.41%)** | **Sum of 20** | **1954 (51.27%)** | **Sum** | **81 (100%)** | **Sum of 20** | **202 (90.58%)** |

Table 2: The twenty most frequent syntactic patterns (ordered by frequency)

Abbreviations' explanation: Abr = Abbreviations; Adj = Adjective; Adp = Adposition; Art = Article; Conj = Conjunction; CPunct = Close Punctuation; Dig = Digits; N = Noun; OPunct = Open Punctuation; Punct = Punctuation; Res = Residual (Foreign Word); V = Verb.

In a previous study [61] we defined two categories of concepts, namely atomic and complex, based on linguistic criteria which were quantified following the morphosyntactic analysis of KOSs. Briefly, atomic concepts are the conceptually undividable units as perceived in the natural language context; into this category also fall the compounds which are terms that can be split morphologically into separate components (words), like the LCSH heading “Hours of labor”, yet the term as a whole represents a single concept. On the other hand, complex concepts are units that are conceptually dividable and comprise two sub-categories. One is ‘enumeration/parataxis’ which recites variant concepts without explicitly declaring any kind of relations between them. Such an example is the DDC term “Volcanoes, earthquakes, thermal waters and gases”. The second sub-category of complex concepts is ‘composites’ which comprise two or more natural language concepts with some kind of relation between them; the LCSH heading “Crimes against peace” is a typical example of a composite concept. Based on the aforementioned categorization of concepts, data derived by the morphosyntactic analysis of each of the conventional KOS reveals the proportion (see Figure 2) of complex concepts within each KOS.

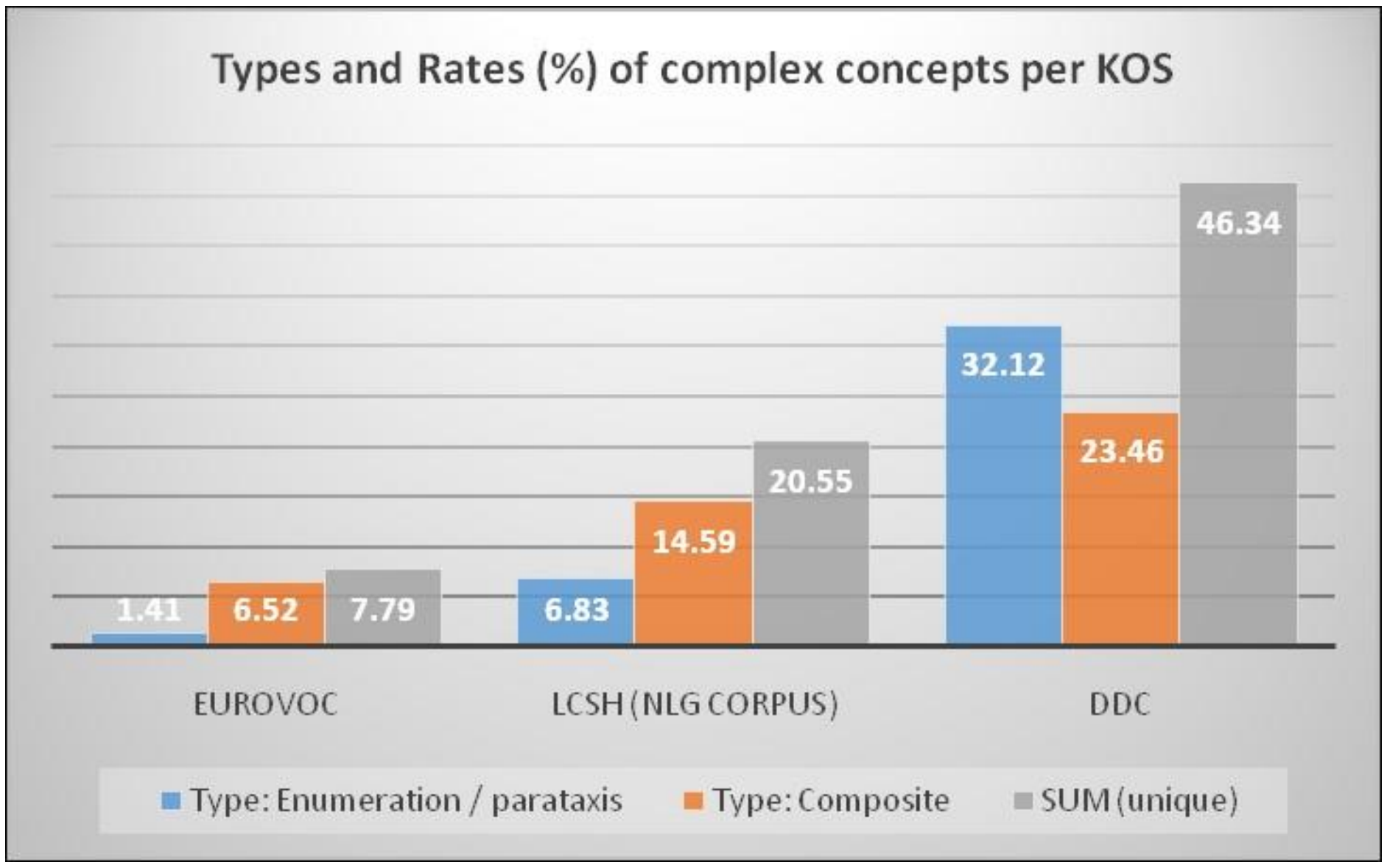

Figure 2: distribution of complex concepts per KOS

At this point, two clarifications seem appropriate. Firstly, Figure 2 does not depict data for CIDOC-CRM because neither CIDOC classes nor CIDOC properties are concepts; therefore, a comparison of this type would result in misleading conclusions. A thorough analysis of this position is presented in the “Ontological approaches: CIDOC-CRM” section of this paper. Secondly, in order to better connect Figure 2 with the results presented earlier in this section, it should be clarified that the word ‘term(s)’ as mentioned in the previous paragraph is used instead of the word ‘entry(ies)’ and refers to each alphanumeric sequence of characters or groups of characters which is defined as an individual listing within the contents of KOSs.

## 3.3. Comparative analysis

Before proceeding with the critique of each KOS examined, a number of generic issues applying to all KOSs are addressed. These issues basically concern relations, not concepts, mainly because concepts are of varying natures, while relations share many characteristics. As already mentioned in the section "KOS as Artificial Language: Concepts and Relations as Functional Units", KOSs define three types of relations: equivalence, hierarchical and associative.

Equivalence relationships manage varying appellations of the same concept or quasi synonyms which form a single concept in the context of the KOS. Despite the fact that these relationships seem quite definitive, they bear a certain degree of ambiguity, since they do not specify whether the cross-reference concerns a synonym, i.e. the relation of a word with an alternative - for example "crazy" *see* "mad"; or a relevant term (concept) - for example "arena" *see* "stadium". Under this rationale, a KOS might not explicitly record whether a reference is assigned from a narrower term to a broader term - for example "cucumber" *see* "fruit vegetable".

The problems of hierarchical relationships are more complicated. The emerging hierarchies are quite generic and the interpretation of the hierarchy is somewhat ambiguous. Three basic types of hierarchies are included in the examined KOSs, though without explicitly mentioning which of these hierarchies applies in each case. The first relationship is the *"is a"* relationship, which is a typical species-to-genus relation. This relationship is based on the inheritance of properties, and its use declares that the concept inherits the properties of the more general one. The second relationship is the "*part of*", which is implemented when a part belongs to a particular whole - for example, the leg is part of the body. The third relationship is the "*instance of*", which applies when an individual instance is related to the class to which it belongs - for example, London is an instance of a city.

These three subcategories of hierarchical relationships could be further divided. Johansson [62] recognizes four different types of the *"is a"* relationship. The first is *subsumption under a genus* which "is the traditional way of creating classificatory trees of natural kinds; in particular, of creating the famous hierarchies of plants and animals in biology". The second, "*Determinable-subsumption*, is not concerned with natural kinds, but with qualities (properties) of different generality. For instance, as a determinate, *scarlet* is subsumed under the determinable *red.*" The other two relationships are *"specification of*" and "*specialization* of"; for example 'painting carefully' is a specification of 'painting' and 'painting a table' is a specialization of 'painting'.

The third category of relationships is the associative, which is the most ambiguous. The relation between two concepts declares a semantic connection which is not further specified in terms of the exact relation between A and B.

Following the analysis of this subsection, it is deduced that, the more explicitly defined, hierarchical relationships are still somewhat ambiguous. Even the taxonomies, which are generated in the LIS domain for categorizing these kinds of relations, constitute evidence of the fuzzy logic that dominates hierarchical relationships [63, fig. 8]. Yet it is a fact that KOSs "use only a small number of paradigmatic semantic relations so far"[63, p.100].

In the following subsections, each dedicated to a particular KOS, a more detailed analysis of the results is presented, along with a comparison and a critique of the datasets studied.

### 3.3.1. Thesaurus: Eurovoc

According to ISO 25964, a thesaurus is a "controlled and structured vocabulary in which concepts are represented by terms, organized so that relations between concepts are made explicit, and preferred terms are accompanied by lead-in entries for synonyms or quasi-synonyms" [64, p.12]. Thesauri have been used in libraries for more than half a century [65] and may vary, both in structure and semantics [66].

At a certain level of abstraction, each term in a thesaurus represents a concept. The term could be a single word or a compound term "that can be split morphologically into separate components" [64, p.3]; for example "copper mines". Although compound terms might be expected to form the exception of the rule "one concept - one word", evidence in Table 1 shows that there is an average of 2.19 words per term in Eurovoc, while Table 2 shows that only 1,364 out of 6,882 Eurovoc terms (i.e. 19.81%) comprised a single noun. Thus, although the number of words in Eurovoc is greater than LCSH, it turns out that LCSH has more complex concepts than Eurovoc, as shown in Figure 2 (20.55% versus 7.79% respectively). Apparently, complex natural language expressions are necessary to express atomic concepts.

Thesauri do not have a build-in mechanism for generating intra-term relations, yet the linguistic analysis conducted in the context of this study did reveal some cases where thesauri use linguistic expressions in order to generate composite concepts. For example, the use of adpositions as depicted in Tables 1 and 2 advocates as regards this practice. The aforementioned evidence, along with the observation that 4.33% of Eurovoc words are Proper Nouns confirm Stellato's observation [67] that, in general, thesauri do not differentiate between classes and instances of objects or between objects and concepts.

Concluding this section, it should be noted that ISO 25964-1 allows for the use of broader terms with and without specification, namely "Broader term (generic)" and "Broader term (instantial)", as well as "Broader term (partitive)" [64, p.13]. However, "there are very few datasets yet that use the ISO hierarchical relation properties" [68, p.41] and Eurovoc is not one of them.

### 3.3.2. Subject Headings: LCSH (NLG corpus)

The subject headings system is the first systematic effort to provide subject access via verbal representation. It constitutes a controlled vocabulary whose main mechanism uses subdivisions in order to further specify the main term which might be a name or a concept. "Library of Congress developed the LCSH for use in its own catalogs. The list was also considered appropriate for the very largest public libraries, some colleges and many university libraries" [69, p.3]. Thus subject access, which was based on verbal indexing during the 20th century both within and outside the USA, was dominated by LCSH [70]. Based on the zeitgeist it was one of the first libraries' vocabularies which was converted into SKOS [71], mainly because both "SKOS and LCSH/MARC have a concept-oriented model" [72, p.24].

LCSH subfields use fewer words than Eurovoc terms for representing concepts (1.99 and 2.19 respectively). However, does not establish a causal nexus that they represent more atomic concepts than Eurovoc terms. Parts of Speech that indicate relations were found to be double the number of those in LCSH; in particular, conjunctions are 4 times as much (3% of the total number of words). An even closer look at the usage of conjunctions reveals a rather inconsistent use. Regarding the usage of "and" in particular, which is largely used as a concept connector, the exact nature of the connection it represents is not specified. In some cases it is used as an intersection (Boolean "and"), as is the case with "Sex and law", "Agriculture and state", "Religion and science", and "Transportation and state". In other cases it is used as a union (Boolean "or"), for example "Brothers and sisters", "Language and languages", "Translating and interpreting" or "Law and legislation".

Unlike thesauri, subject headings bear intra-term relations and have the respective mechanisms to generate them. One such mechanism is the use of punctuation. A more detailed look at the data of this study shows that the usage of punctuation is extensive within subfields since six (6) out of twenty (20) patterns contain punctuation, such as parentheses. Based on data depicted in Table 2, the fifth pattern, covering a percentage of 3.58%, is [N + OPunct + N + CPunct], where OPunct is the opening of a parenthesis and CPunct is the closure. In terms of semiotics, parentheses are used to either define the context within which the term is used or to further specify the preceding entity. Such examples are the headings "Cooking (Apricots)" and "Impediments to marriage (Canon law)", where there is an intra-term paradigmatic relationship. In addition, "*LCSH* has methods other than the use of parenthetical qualification to fix the referential semantics of such terms: for instance, it could be embedded in a 'In' type heading, e.g., 'Flight in birds'; or established as part of a phrase heading, e.g. 'Bird flight'; or qualified by a subheading, e.g., Birds -- Flight" [18, p.21]. It is observed that all the above have an effect on alphabetic sorting, which seems quite important, even for the present time. This importance is particularly recorded, among others, in the use of the third most frequent syntactic pattern, namely "N + Punct + Adj", where Punct dictates the use of a comma value. The noun-adjective inversion is implemented so that the heading is alphabetically sorted under the noun.

Table 1 and Table 2 show that there is a noticeable use of adpositions in LCSH, holding the highest distribution among the conventional KOSs. This specific type of POS is used to associate nouns comprising a composite concept; for example, *offenses against the person* or *exercise for youth*.

Another observation is that headings very similar from a linguistic perspective have significant semantic differences. A representative example of this case is the headings

“Philosophy and religion” and “Religion -- Philosophy”. As stated in the scope note of the second authority “Here are entered works on the philosophy of religion. Works on the reciprocal relationship and influence between philosophy and religion are entered under Philosophy and religion”.

The aforementioned remarks emphasize the efforts of formalization which focuses on the value of the heading - as a string of characters - in order to express variant semantics. It is evident that this practice asks for tacit knowledge as a prerequisite in order to be interpreted, while there is no method or mechanism of explicitly declaring relations between the components. Instead, this can be achieved by combining values in any way considered appropriate. Therefore, the aphorism of Kwaśnik that “LCSH is a mishmash of terms and relationships that was built up incrementally over the course of a century” [73, p.21] does not seem too far out.

### 3.3.3. Classification schemes: DDC

Classification systems were developed and used by libraries for managing the shelving of physical objects, mainly referring to the books of a collection. Later they evolved into multi-functional tools and, currently, they are used even in the context of digital libraries where physical objects and shelving are not applicable. Among classification schemes, DDC is very popular, used in more than 135 countries, and has been translated into more than thirty languages according to OCLC’s data for 2017. In order to be usable in the context of the semantic web, DDC is also provided as linked data, and the effort of assigning each number to a different URI dates back many years [74].

DDC’s basic mechanism is to generate a hierarchical tree structure using notation. The Notational hierarchy, as stated in the DDC introduction, “is expressed by length of notation. Numbers at any given level are usually subordinate to a class whose notation is one digit shorter” and “[s]tructural hierarchy means that all topics (aside from the ten main classes) are part of all the broader topics above them. The corollary is also true: whatever is true of the whole is true of the parts”. This chain creates a mono-hierarchical relationship where each class can have only one broader class.[5]

According to DDC’s introduction (principle 4.14), “Since the parts of the DDC are arranged by discipline, not subject, a subject may appear in more than one class”. Like all pre-coordinate systems, DDC is defined by the a-priori choice of a dominant category under which all potential subdivisions are included. The problems seem to increase when a document has to be assigned a dominant category which is provided as a subcategory within the system. As Hjørland states on this topic “It is my claim that this type of analysis, which determines the priorities of the viewpoints to be taken on a document, is not optimal in every situation. One can imagine researchers working on technical aspects of the election process who wish to compare them in several countries. For such a person the election would be the central subject, and it would be inconvenient if this were a subtopic of History and India" [75, p.178].

[5] Certainly, this is not unconditional, since there are exceptions like interdisciplinary numbers.

Moreover, Green & Panzer [76] examined the types of hierarchical relations within the DDC classes and identified several types of them - for example, whole-part or specialization - for which no certain formalism was implemented. In addition, Mazzochi reports that "perspective hierarchies, which are used by classifications systems such as the Dewey Decimal Classification, function more contingently. Usually, they are not provided with the same logical properties of generic hierarchies" [77, p.370]. Conclusively, there is no strong mechanism/methodology in DDC concerning how classes (or the meaningful parts of classes) are bound together.

It is expected that a notation system would present limited dependence on verbal representations, which is mostly true in terms of the formalism of lexical representations. DDC maintains a distance from the other KOSs concerning the variety of identified patterns, since it counts one pattern for every 3.38 entries, while its twenty most frequent patterns barely exceed 50% of the number of total entries. The percentage of entries covering the twenty most frequent patterns reaches 85.41% in LCSH and 88.17% in Eurovoc. According to Mitchel et al [78, p.94] the notation (i.e. the DDC number), the lexical representation of the notation and the URI of a DDC class may be used interchangeably as long as they are *nomens* of the same *thema*. Therefore, a more detailed view of linguistic analysis of the lexical representations of the *themas* may provide more information about the *thema* itself.

Table 1 shows that DDC has the lowest rate among the three traditional KOSs of POS assigned to entities, and the highest proportion assigned to relations; almost 16% of its words indicate relations. This observation, along with the fact that DDC comes first among the studied KOSs concerning the assignment of POS-entities (1.98 per entry), it shows that there are more complex concepts in DDC than in other KOSs. These concepts could be parts of a generic category (for example "025.39 Recataloging, reclassification, re-indexing") or more complex relations between concepts (such as 302.5 "Relation of individual to society"); or a Boolean intersection (for example "401.4 Language and communication"). Specifically, the use of "and" in DDC is similar to the use of "and" in LCSH, reminding the difficulty identified earlier in distinguishing between the cases in which it is used as a union and the ones where it is used as an intersection. In DDC even more complex cases are identified, as in the case of "378.12 Faculty and teaching", where the use of "and" cannot be accurately defined as a Boolean operator, since in the scope note it is stated that "Standard subdivisions are added for faculty and teaching, for faculty alone".

Table 2 shows that the fourth most frequent pattern contains digits which correspond mainly to chronological information. DDC often uses sporadic years or other time intervals, predefined or not, as in number 943.08 the value "1866-" which indicates the History of Germany from 1866 onwards.

### 3.3.4. Ontological approaches: CIDOC-CRM

Poli and Obrst notice that the notion of ontology "comes with two perspectives: one traditionally from philosophy and one more recently from computer science. The philosophical perspective of ontology focuses on categorial analysis […]. Prima facie, the intention of categorial analysis is to inventory reality. The computer science perspective of ontology, i.e. ontology as technology, focuses on those same questions but the intention is distinct: to create engineering models of reality, artifacts which can be used by software, and perhaps directly interpreted and reasoned over by special software called inference engines, to imbue software with human level semantics" [79, p.1]. LIS theory is more familiar with the philosophical perspective, while LIS practice uses the computer science perspective. This paradox causes some inconsistencies in the LIS domain.

In this study we follow Gruber's classical definition where an "ontology is an explicit specification of a conceptualisation. The term borrowed from philosophy, where Ontology is a systematic account of Existence. For AI systems, what 'exists' is that can be represented" [80, p.908]. So an "ontologist moves away from questions about reality itself and turns to questions about the representation of reality" [81, p.88]. It is evident that "the development and application context offered by linked data has helped to transform the perception of ontologies from the complex and stratified knowledge systems grounded in philosophical theory to the more pragmatic tools used to stitch together bits of information to create the fabric of a Web of linked data" [82, p.266].

CIDOC-CRM is a high level ontology which aims to integrate information for cultural heritage data [83]. Its inclusion in this study is not based on its essence or functionality as a mainstream KOS but to serve as a criterion for comparing KOSs with ontologies. CIDOC-CRM, as a typical ontology, is divided into two main parts: classes and properties, where a class is a "category of items that share one or more properties" [84, p.2] and a property is a "defining characteristic that serves to define a relationship of a specific kind between two classes" [84, p.4]. The fundamental mechanism of ontologies in general, as well as of CIDOC-CRM in particular, is the correlation of two *classes* (Domain and Range) through a *property*. Classes and properties are studied here as individual groups (datasets) in order to demonstrate the variant POS used in each of them and, through evidence, to demonstrate how dissimilar the modelling of an ontology is compared to traditional KOSs.

Concerning the linguistic analysis, first it should be clarified that the ontology does not share the same characteristics as the examined KOSs. According to SKOS, every skos:Concept is an instance of owl:Class. The distinction between ontologies and KOSs is described through SKOS by Jupp et al, where they note that "SKOS is not a language for modelling ontologies, the data model itself is in fact described as an OWL ontology. A particular KOS is then represented as an instantiation of this ontology" [85, p.2]. Consequently, KOSs are value vocabularies and the words analyzed are the values. Labels of classes and properties of the ontology are not of this kind. As mentioned before, the basic reason for linguistically analyzing the ontology had to do with designating the difference in linguistic characteristics present in properties, since the novelty in ontologies are the properties, and not the hierarchy which might or might not exist[6].

Table 1 shows that, within CIDOC-CRM classes, POSs assigned to relations are almost absent. On the contrary, the majority of words categorized as relations (66.73%) have been derived from the *properties* part of CIDOC. The evidence of this study shows that this is the only case where verbs play a central role; practically every property contains a verb. In traditional KOSs verbs are hardly present, while, in ontologies, there are more properties (i.e. relations) than classes.

## 3.4. Critique

In the LIS domain the categorization of relationships as paradigmatic and syntagmatic is quite popular. As already mentioned, the latter constitutes an open class that does not allow the explicit definition of relations between concepts. When it comes to the modelling of ontologies, the aforementioned categorization is reconsidered, since all relations are defined explicitly. In computational processing, as perceived and implemented in the context of the semantic web, concepts are determined by their properties, and not by their lemma definition. On the contrary, scope notes have an important role in the improvement of the human indexers' understanding [86]. In essence, RDF properties are relations. Therefore, in the context of the semantic web, the decision has been made: the appellation has nothing to do with reasoning; except from cases of specific sorts of values such as dates or any other unit of measurement; this specific sort of data corresponds to the Datatype properties of OWL. SKOS URI is a nomen which, in contrast to the *nomens* of the type 'control access names', is not meant to be managed by humans. Indeed it may be the only *nomen* that indicates concepts unambiguously.

The basic mechanism "DomainClass-hasProperty-RangeClass" of ontologies is very close to the main construction "Subject-Verb-Object" (SVO) of natural language. As Hoeppe puts it, "Setting out to study representation as a verb (an activity), and not only as a noun (its end product), has been one of the most fruitful moves in the social studies of science and technology" [87, p.1078]. This approach creates the space for "a consistent conceptual structure that reveals interaction among their elements" [88, p.179]. The expressiveness of ontologies is enhanced due to the fact that they allow for explicit declarations of restrictions upon relations. Therefore, their expressiveness could be paralleled to the natural language syntax in its primitive form. On the other hand, ontologies are more precise than natural language (meaning that they are free from the inherent ambiguity of natural language) and, by nature, they are machine-processable. Therefore, in a hypothetical *expressiveness pyramid* of the KOSs studied, ontologies would be placed on top since they can be more expressive than conventional KOSs.

[6] It is acknowledged that in OWL everything is a subclass of Thing, but this does not mean that every class of the ontology follows an internal hierarchy.

Among traditional KOSs, the closest to ontologies are thesauri because they have a rather coherent definition of what a concept is and they assign specific relations between concepts. But thesauri relations are often limited to generic associative ones (the informational value of which is limited) and to hierarchical ones. However, although a hierarchy provides information on certain things and is highly important for cognition [60, Chapter 7], it alone cannot lead to adequate inference. As discussed in section 3.3.1., atomic concepts in conjunction with the limited number of relations result in thesauri having relatively restricted expressiveness. On the other hand, the high level of formalism allows them to be machine-processable, albeit this machine processability can be fruitful at a basic level of inference but not for reasoning. Thus, thesauri should not be considered ontologies because, as noted by Kless et al, "thesauri and ontologies need to be treated as 2 orthogonal kinds of models with superficially similar structures" [89, p.1348].

On the other hand, LCSH is highly expressive since many kinds of relations lay among and within the fields and the subfields. As shown in subsection 3.3.2., when paradigmatic relationships between functional parts are not adequate, natural language emerges to provide a solution. In this way many composite concepts are generated, as depicted in Figure 2. A representative example is the heading "Turning water into wine at the wedding at Cana (Miracle) in art" (http://id.loc.gov/authorities/subjects/sh97001686). But LCSH focuses on value (string) formalism and not on semantics or structural formalism. The flexibility of generating relations based on natural language is not machine-processable; at least it does not bear the kind of processability required in the context of the semantic web. Since headings incorporate a great number of concepts through both the association of subfields and intra-term relations, inference could lead to misleading associations. A representative example of this problematic inference has been illustrated by Spero who, by following the LCSH relations, shows the connection between the heading "Doorbells" and the heading "Mammals" entitling their work "LCSH is to Thesaurus as Doorbell is to Mammal: Visualizing Structural Problems in the Library of Congress Subject Headings" [90]. The problematic inference is generated because concepts are not defined by their properties, i.e. the RDF/SKOS relations.

Finally, classification schemes represent knowledge in a completely different way, mostly due to the fact that they are not concept-centric but discipline-centric. Concerning the DDC notation, the rationale behind it is that it is a condensed literal; for example, "*190*" stands for *"Modern Western & other noneastern philosophy".* As shown in Figure 2, almost half of the terms in DDC are complex, which are realized through various structures. The most popular category is parataxis, which exceeds one third of terms. A common practice to formulate this category is the use of words like "other" or "etc" which signifies open classes. Such structures are understandable by the human mind, thus boosting expressiveness. Yet, as in LSCH, they do not constitute *machine-processable information.*

Consequently, selecting a knowledge organization or representation scheme directly affects the potential of expressiveness as well as the machine processability. To elaborate on this statement let us consider the term "Violence of students towards teachers" which features three elements, namely "students", "teachers" and "violence". In a post-coordinate system where there is no indication of the direction of relations, the intersection of the concepts could not indicate who the subject of the violence was, i.e. whether it was exercised from teachers to students or vice versa. On the other hand, pre-coordinate systems, such as Subject headings or DDC, would sufficiently represent such a subject, most likely by using an ad hoc solution which would lack structural formalism. An ontology could accurately represent it with two classes and a relation (property), specifying the direction of the relation. This is very crucial because the "rigorous definitions to characterize formal relations will be a major step toward enabling information scientists to achieve interoperability among ontologies in support of automated reasoning across data derived from multiple domains" [91, p.222].

# 4. Conclusions

This study has revealed that, in the context of traditional library-based KOSs, there is no clear or consistent definition of what constitutes a concept. The morphosyntactic analysis indicated that there is no clear distinction between named entities and concepts; proper nouns appear along with common nouns. This means that there is no differentiation such as the one defined between the classes and instances of ontologies. This type of analysis also indicated that various types of intra-term relations are identified both in LCSH and DDC, but there is no built-in mechanism or methodology to define them explicitly. Only the position in the hierarchy, or sometimes a qualifier, defines the context in which the concept acquires its meaning. Even so, the way the hierarchy is structured may lead to poor semantics since it is not always based on "*is a"* relationships.

Especially in the case of LCSH, the emphasis is placed on the structure of terms as strings which formalize the value of the heading in order to create an appropriate identifier for string matching. Inter-term relations come second in order of importance. The LCSH construction mechanism manages semantics very thinly in terms of its formalistic representations, while using natural language extensively for enhancing its flexibility and expressiveness; however, this expressiveness cannot be transferred to the Semantic Web. Additionally, LCSH seems to take for granted a significant amount of latent information but it is well known to the people who work on data modelling that "what looked simplest […] turned out to be diabolically hard to model" [92, p.7]. The frequent use of enumeration in DDC (as depicted through the NLP approach) renders the scheme not that concept-centric, while the hierarchies produced are also weak in terms of semantics. Instead, the semantics of relations in thesauri are more definitive. In addition, due to the explicit semantics of concepts, more relationships could be implemented - beyond hierarchical ones - similar to the ones used in ontologies. For example, a statement such as "Deer eats grass" can be expressed in a thesaurus using two concepts and a predicate with pre-defined direction. By contrast, in LCSH and DDC such statements can be, and are, expressed as complex concepts.

Current trends focus on the modelling of relationships since this is a prerequisite for the reasoning process which results in the assignment of meaning to concepts. In conventional KOSs the modelling of relations falls short, while natural language is used for defining intra-term relations which allow for the creation of complex concepts. But relations are very difficult to model as long as there is a lack of atomic concepts. The more complex the concept is the harder it becomes to produce precise relations between the two; sometimes it is not possible at all. This lack of precision in LCSH can lead to door bells being related with mammals, as mentioned in section 3.4. As Wilmont et al state, "A relation describes the behaviour between two or more concepts. [...] As is the case with concepts, a relation also does not become meaningful without a thorough, concrete understanding of the concepts it binds. Hence, concepts and relations depend on each other to acquire meaning" [93, p.77]. This way a meaningful network is created where the epicentre of the concepts' interpretation is set on their relations. Within a semantic network, with nodes and arcs (edges), it is absolutely clear that each node is joined with another node through an arc, which represents a relationship, i.e. property, which (as revealed through our study of CIDOC-CRM) is always expressed with a verb.

As Zeng and Mayr note in an article about the use of KOS as linked open data (LOD), there is still a long way to go to make KOS be recognized as knowledge bases and semantic tools in the context of the semantic web [94]. What is stressed in our study is that, in order for this to be achieved, the issue is not only about converting data from one format to a newer one; this process alone may hold us back regarding the exploitation of the full potential of the new format. This choice is like using a smart-phone only as a telephone device, i.e. use it only for making phone calls. The main issue is to embrace a different rationale overall. "Re-engineering KOS by ontological principles generally requires a new way of thinking" [95, p.67] and the re-engineering of KOSs demands re-conceptualization. Thus, the point is not just to express a conventional KOS in new languages (such as OWL), but to build on the new possibilities emerging through new theoretical and technological infrastructure in order to achieve more accurate and sophisticated expressions of its meaning. Ultimately, paraphrasing the famous proposition of Wittgenstein "[t]he limits of my language mean the limits of my world" [96, para. 5.6], we could say that the limits of our artificial languages mean the limits of our Knowledge Organization Systems.